\documentclass[reprint]{JASA}

\usepackage{siunitx}
\usepackage{todonotes}

\let\added\undefined      
\let\deleted\undefined    
\let\replaced\undefined
\let\listofchanges\undefined
\usepackage[defaultcolor=red, final]{changes}
\newcommand{\revcolor}{black}

\usepackage[opacity=1]{background} 
\SetBgContents{\parbox{17cm}{The following article has been submitted to the Journal of the Acoustical Society of America.\\After it is published, it will be found at  \url{http://asa.scitation.org/journal/jas}.}}
\SetBgScale{1}
\SetBgAngle{0}
\SetBgPosition{current page.north east}
\SetBgHshift{-6.0cm}
\SetBgVshift{-0.4cm}

\newcommand{\setcomplex}{\mathbb{C}}
\newcommand{\setpsd}{\mathcal{S}_M^+}
\newcommand{\setC}{\mathcal{C}}
\newcommand{\zeros}{\mathbf{0}}
\newcommand{\ones}{\mathbf{1}}

\newcommand{\bx}{\mathbf{x}}
\newcommand{\by}{\mathbf{y}}
\newcommand{\bu}{\mathbf{u}}

\newcommand{\bw}{\mathbf{w}}
\newcommand{\balpha}{\bm{\alpha}}
\newcommand{\bSigma}{\bm{\Sigma}}
\newcommand{\bGamma}{\bm{\Gamma}}
\newcommand{\bC}{\mathbf{C}}

\newcommand{\bZ}{\mathbf{Z}}
\newcommand{\bM}{\mathbf{M}}
\newcommand{\mId}{\mathbf{I}}
\DeclareMathOperator{\sinc}{sinc}
\DeclareMathOperator{\diag}{diag}
\DeclareMathOperator{\rank}{rank}
\newcommand{\argmin}{\operatornamewithlimits{arg min}}
\newcommand{\herm}{\mathsf{H}}

\newcommand{\frob}{\mathsf{F}}
\newcommand{\nucl}{*}
\newcommand{\indic}{I_{\setC}}
\DeclareMathOperator{\fpenalty}{\Omega}

\DeclareMathOperator{\prox}{\textbf{prox}}

\DeclareMathOperator{\expectancy}{\mathrm{E}}
\newcommand{\lsloss}{l}

\newcommand{\vsense}{\by}
\newcommand{\vfield}{\bx}
\newcommand{\vnoise}{{\bw}}
\newcommand{\vgains}{\balpha}
\newcommand{\vgainsest}{\tilde{\vgains}}
\newcommand{\vgainsestscaled}{\vgainsest_{\text{sc}}}

\newcommand{\vgainsref}{\vgains_{0}}
\newcommand{\mgains}{\bC}
\newcommand{\mgainsest}{\widetilde{\bC}}
\newcommand{\mcovfieldmodel}{\bSigma_x}

\newcommand{\mcovsensemodel}{\bSigma_y}
\newcommand{\mcovsensesample}{\widetilde{\bSigma}_y}

\newcommand{\mcohsensesample}{\widetilde{\bGamma}_y}

\newcommand{\sgainmean}{\overline{\alpha}}
\newcommand{\varfield}{\sigma_x^2}
\newcommand{\varfieldi}{\sigma_x^{-2}}

\newcommand{\varnoise}{\sigma_w^2}
\newcommand{\wavelength}{\lambda}
\newcommand{\soundspeed}{c_0}
\newcommand{\dist}{d}
\newcommand{\regp}{\nu}
\newcommand{\stepgrad}{\gamma}

\newcommand{\nbsensors}{M}
\newcommand{\nbsnap}{L}
\newcommand{\pbzero}{$\mathcal{P}_0~$}
\newcommand{\pbone}{$\mathcal{P}_1~$}
\newcommand{\pbi}{$\mathcal{P}_i~$}

\begin{document}

\title[Sensor gain and phase calibration in the ambient noise]{Gain and phase calibration of sensor arrays from ambient noise by cross-spectral measurements fitting}
\author{Charles Vanwynsberghe}
\email{charles.vanwynsberghe@tii.ae}
\affiliation{Technology Innovation Institute, 9639 Masdar City, Abu Dhabi, United Arab Emirates}
\author{Simon Bouley}
\affiliation{MicrodB, F-69134, Écully, France}
\author{Jérôme Antoni}
\affiliation{Univ Lyon, INSA Lyon, LVA, EA677, 69621, Villeurbanne, France}

\preprint{Author, JASA}		

\date{\today}

\begin{abstract}
We address the problem of \textit{blind} gain and phase calibration of a sensor array from ambient noise. The key motivation is to ease the calibration process by avoiding a complex procedure setup. We show that computing the sample covariance matrix in a \textit{diffuse field} is sufficient to recover the complex gains. To do so, we formulate a non-convex least-square problem based on sample and model covariances. We propose to obtain a solution by low-rank matrix approximation, and two efficient proximal algorithms are derived accordingly. The first one solves the problem modified with a convex relaxation to guarantee that the solution is a global minimizer, and the second one directly solves the initial non-convex problem. We investigate the efficiency of the proposed algorithms by both numerical and experimental results according to different sensing configurations. These show that efficient calibration highly depends on how the measurements are correlated. That is, estimation is achieved more accurately when the field is spatially over-sampled.
\end{abstract}


\maketitle


\section{Introduction}
\label{sec:intro}
Calibration refers to the task of fixing the gain of a sensor accurately. The task is conventionally done on a single element, but can also be performed on a distributed network \cite{Delaine2019,BarceloOrdinas2019} or a synchronous array. The latter configuration is considered in this paper. Several acoustical problems rely on the use of sensor arrays (\textit{e.g.} source localization and identification) and are therefore sensitive to potential model uncertainties \cite{Gilquin2019}. Both amplitudes and phases of sensor gains are a part of such critical uncertainties \cite{Friedlander1993,Weiss1990, Chen2013,Cui2017}.

Classically, gain calibration is done on a single element either by the reciprocity or  the comparison method \cite[ch. 24]{Rossing2014}. The reciprocity method is adopted in standardized protocols to calibrate reference microphones \cite{Frederiksen2013}. In spite of its reliability, going for an element-wise calibration with a large sensor array is cumbersome, if not infeasible in practice. Consequently, attempts were made to calibrate the whole array in one shot. The overall picture is as follows in the noiseless case: provided $1$ sensor reading, one needs to discriminate the (unknown) sensor gain from the (unknown) actual value of the physical field itself, resulting in $2$ unknown parameters for $1$ reading. By extension, $\nbsensors$ sensor readings correspond to $2\nbsensors$ unknowns, so the problem is under-determined for any array of $\nbsensors$ sensors. However, reducing the number of unknowns can be tackled in different ways. A first example is to rely on the range space deficiency in which the physical field lies \cite{Balzano2008,Balzano2007}. A sequence of well-calibrated measurements is then supposed to lie in a signal subspace whose dimension is small compared with $\nbsensors$. A second example rather assumes that the physical field is generated by a small number of sources, and is based on sparse approximation \cite{Bilen2014,Gribonval2012}. Moreover, if sensor gains are somehow correlated by the underlying physical phenomena (\textit{e.g.} temperature), they can also be constrained into a subspace \cite{Ling2018,Ling2015} (\textit{i.e.} of low dimension as compared to $\nbsensors$).

One context that has rarely been explored in acoustics is \textit{in situ} calibration with an ambient noise physical field. It is of potential interest to several applications, such as in underwater acoustics (\textit{e.g.} source monitoring \cite{Menon2012a}, passive fathometer \cite{Siderius2010}, geoacoustic inversion \cite{Yardim2014}, Green function retrieval \cite{Roux2005}, thermocline \cite{Li2019,Godin2010} or temperature \cite{Woolfe2015,Weaver2001} estimation), in seismic sensing (\textit{e.g.} tomography \cite{Shapiro2005}, detection \cite{Seydoux2016}), or structural health monitoring by tomography \cite{Druet2019}. Other inverse problems applied to imaging are listed in \cite{Garnier2016}. This non-exhaustive list depicts the diversity of applications where sensor gain uncertainty surely causes estimation issues and, therefore, should be mitigated.

Rare works port the attention on how to benefit from ambient noise to calibrate sensors. In \cite{Pavlis1994}, a seismometer is calibrated with ground noise, and placed in proximity to a supplementary seismometer of reference whose absolute gain is known. This is the  aforementioned comparison method: it relies on the calculus of the transfer function between both seismometers. It assumes that the latter record the same physical field, and it is a sensor-wise procedure. More recent efforts were made by Abkar \textit{et al.} to obtain the microphone gains \cite{Akbar2021,Akbar2020} of a spherical array, with the help of a diffuse field. These authors emphasize the practicality of this approach because it is easier to generate a spatially scattered acoustic noise than a controlled source (or more) in free field conditions. However, the suggested method is energy-based, and provides the amplitude part of the gain only.

We propose a sensor gain and phase calibration where the ``ambient noise" precisely refers to a \textit{diffuse field}. In such conditions, cross-spectral densities between sensors can be expressed by a well-known second-order statistic model \cite{Nelisse1997,Jacobsen1979}. When the cross-spectral readings are arranged in the form of a covariance matrix, we show that the problem can be turned into a low-rank matrix approximation problem\cite{Davenport2016}. To the best of our knowledge, the solution proposed in this paper is distinct from the state-of-the art by the fact that it relies on second-order statistical measurements (the sample covariance matrix) rather than on a collection of snapshot vectors. As a consequence, the complexity is independent of the observation time duration and, due to the averaging of noise, good performance is reached even with low signal-to-noise ratios. Moreover, the low-rank matrix approximation problem is often addressed in contexts where dimensions are large (\textit{e.g.} in machine learning \cite{Davenport2016}), and efficient methods were proposed accordingly. Leveraging low-complexity methods is relevant here, because it allows the simultaneous calibration of a large number of sensors.

The paper is organized as follows. Section \ref{sec:model} describes the sensing model based on the diffuse field assumption. The trivial derivation of a gain estimator is presented with its main limitations for a practical use in Sec. \ref{sec:vanilla}, \replaced{and will be used as a benchmark for the performance analysis in experiments}{which will serve as the baseline in experiments}. Then, Sec. \ref{sec:sensor_calibration} defines the least-square problem in the original non-convex form, followed by a convex relaxation to ensure that the obtained solution is the global minimizer. Since the estimated gains are relative up to a constant, Sec. \ref{subsec:rel_to_abs} discusses different options for relative to absolute value conversion. Finally, two efficient proximal algorithms \cite{Parikh2014} that solve the non-convex and convex problems are derived in Sec. \ref{sec:solvers}. The performance of these solvers with regard to signal-to-noise ratio, acquisition time, model mismatch and the density of spatial sampling are studied via simulations in Sec. \ref{sec:numerical_experiment}. At last, an experimental validation is presented in an acoustic reverberant environment to calibrate a $43$-microphone array in Sec. \ref{sec:experimental_results}. A final conclusion summarizes the whole results and suggests relevant practices to perform calibration in ambient noise.

\subsection*{Code and Github}
A Python implementation of the described algorithms is available online at \url{https://github.com/cvanwynsberghe/sgcal-jasa}.

\subsection*{Mathematical notations}

The notations in the paper are as follows. Bold lowercase (resp. uppercase) letters describe vectors (resp. matrices). The identity matrix of size $N \times N$ is defined by $\mId_N$; $\zeros_N$ stands for the column vector of $N$ zeros, and $\ones_{M \times N}$ stands for the $M \times N$ matrix of ones. $\diag(\bm{v})$ is the diagonal square matrix whose diagonal contains the elements of the vector $\bm{v}$. $\odot$ and $\oslash$ respectively stands for the element-wise product and \added{element-wise} division of two equal-size matrices. $\| \bM \|_\frob$ denotes the Frobenius norm of the matrix $\bM$\added{, and $\|\bM\|_2$ denotes the maximal singular value of $\bM$}. $\bM^\herm$ stands for the \added{conjugate transpose} of matrix $\bM$.

\section{Covariance-based sensing model in ambient noise}
\label{sec:model}

Throughout this paper, the considered scenario involves an array of $\nbsensors$ sensors providing pointwise and omnidirectional measurements in a homogeneous media, meaning that the propagation speed $\soundspeed$ is constant. One given frequency $f$ relates to one wavelength value, denoted $\wavelength$. The sensing model is expressed in the frequency domain after applying the Discrete Fourier Transform (DFT) on a collection of time series to obtain $\nbsnap$ snapshots indexed by $l$.

This paper considers the sensor calibration problem for which an unknown frequency-dependent multiplicative and complex-valued gain affects the output value of each sensor. In the noisy case, the considered model for the $l$-th snapshot at frequency $f$ reads
\begin{equation}
    \label{eq:model_snap}
    \vsense_l = \diag(\vgains) \vfield_l + \vnoise_l,
\end{equation}
where $\vgains \in \setcomplex^{\nbsensors \times 1}$ denotes the set of unknown sensor gains, $\vsense_l \in \setcomplex^{\nbsensors \times 1}$ are the values read by the sensors, $\vfield_l$ are the true values of the field, and $\vnoise_l$ denotes the additive sensor noise (frequency dependence will be systematically omitted for ease of notation). By considering complex gains, both amplitude and phase variations are taken into account.

The physical field and the sensor noise realizations are supposed to be two independent, zero-mean, \replaced{ergodic}{random} processes. \replaced{The physical field is generated by a spatially isotropic ambient noise; it is defined by the covariance matrix}{The covariance matrix of the physical field is } $\expectancy\{ \vfield_l \vfield_l^\herm \} = \varfield \mcovfieldmodel$, for any $l$, where $\varfield$ denotes the variance of the field and $\mcovfieldmodel \in \setcomplex^{\nbsensors \times \nbsensors}$ contains the coherences between sensors. Similarly, the covariance matrix of sensor noise is $\varnoise \mId_\nbsensors$, which reflects independent variables, each with the same variance $\varnoise$.

When the physical field is ambient noise\footnote{From this point, we strictly distinguish \textit{ambient noise} (characterizing the physical field) and \textit{sensor noise} (\textit{i.e.} the additive non-propagative noise which originates from the acquisition device) in order to avoid any ambiguity between both.}, a well-established model consists in considering that the field is the integral of uncorrelated plane waves of homogeneous variance over the sphere. The latter is known as \textit{diffuse field} \cite{Jacobsen1979}. If  uncorrelated plane waves have variance $\varfield$ over the whole sphere, it reads \cite{Hill1995}:
\begin{equation}
    \label{eq:model_cov}
    [\mcovfieldmodel]_{mn} =  \sinc\left( \dfrac{2\pi}{\wavelength} \dist_{mn} \right),
\end{equation}
where $\dist_{mn}$ is the distance between the $m$-th and the $n$-th sensors. Also, the covariance matrix from the sensor readings can be derived according to the snapshot model in Eq. \eqref{eq:model_snap}:
\begin{equation}
    \label{eq:sample_cov_total}
    \mcovsensemodel = \expectancy\{\vsense_l \vsense_l^\herm\} =\varfield \diag(\vgains) \mcovfieldmodel \diag(\vgains)^\herm + \varnoise\mId_\nbsensors.
\end{equation}
\added{In practice, the advantage of this model is to depend on the relative positions of the sensors rather than the absolute positions: the values of $\dist_{mn}$ should be precisely known as long as the the array structure is well-known.}

\replaced{The}{This} covariance {\color{\revcolor} $\mcovsensemodel$} can be approximated by the measurements from the zero-mean sensor readings. The sample covariance matrix, denoted $\mcovsensesample$, is estimated from the $\nbsnap$ snapshots as
\begin{equation}
    \label{eq:sample_cov_def}
    \mcovsensesample = \dfrac{1}{\nbsnap} \sum_{l=1}^{\nbsnap} \vsense_l \vsense_l^\herm.
\end{equation}
We emphasize that the use of DFT to build the snapshots $\vsense_l$ is impacting and should be well-tuned. Indeed, $\mcovsensemodel$ describes the diffuse field for the frequency $f$, whereas $\mcovsensesample$ results from DFT snapshots, and the DFT \replaced{is equivalent to}{acts as} \replaced{a bank of  bandpass}{low-pass} filters having the same \replaced{resolution}{precision} frequency denoted $\Delta f$. \replaced{The effect of each bandpass filter is to integrate the spectrum}{This results into an integral} over a frequency interval of width $\Delta f$, \replaced{however the integration}{that} is not taken into account in the derived model \eqref{eq:model_cov}. Keeping consistency between the model $\mcovsensemodel$ and the measurement $\mcovsensesample$ requires that DFT is sufficiently narrowband \cite{Nelisse1997}. Then, respecting
\begin{equation}
\label{eq:condition_dft}
\Delta f \ll \dfrac{\soundspeed}{\dist_{mn}} \:  \forall \: m, n
\end{equation}
is a necessary condition to preserve the validity of the model \eqref{eq:model_cov}. \added{In practice, the value of $\Delta f$ can be controlled by the number of time samples at the input, the taper function and the sampling frequency}. In the acoustical context, this condition should be easily met.

An interesting aspect of using the sample covariance matrix rather than snapshots is to reduce computation complexity and memory usage: while the amount of data to process equals $\nbsnap\nbsensors$ and scales linearly with the number of snapshots, it remains constant (and equals $\nbsensors^2$) with the sample covariance matrix.

\section{Vanilla solution under strong conditions}
\label{sec:vanilla}

In this section, we show that a straightforward solution can be derived under the noiseless assumption. If $\varnoise \ll \varfield$, plugging in the sample covariance matrix in Eq. \eqref{eq:sample_cov_total} rewritten as an element-wise matrix product, results in the relation $\mcovsensesample \approx \varfield {\color{\revcolor}\mgains} \odot \mcovfieldmodel$ \added{where we define the matrix $\mgains=\vgains\vgains^\herm$}. Then, a simple vanilla estimation of $\vgains$ consists in obtaining the principal eigenvector of the matrix $\mcovsensesample \oslash \mcovfieldmodel$. This solution provides relative gains, \textit{i.e.} up to a complex constant -- this point will be discussed in detail in Sec. \ref{subsec:rel_to_abs}.

The idea is simple, but has a caveat: it is unstable in practice because of the element-wise division. Instability occurs if the coefficient $[\mcovfieldmodel]_{mn}$ is small for some given index pair $(m, n)$, so that the presence of an additive error term in $[\mcovsensesample]_{mn}$ is magnified by the division operation. This case exists for example if $\varnoise \neq 0$, \textit{i.e.} even with low sensor noise. The same issue arises if a small additive error term exists in $[\mcovfieldmodel]_{mn}$, \textit{i.e.} in the model itself. Thus, there is potentially a high sensitivity to sensor noise, but also to model mismatch.

A simple rule of thumb to safely use the vanilla approach is to keep
\begin{equation}
    \label{eq:condition_lowfreq}
    \wavelength \gg \dist_{mn} \; \text{for} \;  1 \leq m, n \leq \nbsensors.
\end{equation}
This condition prevents the $\sinc$ function in \eqref{eq:model_cov} from being close to zero, and leads to the approximation
$\mcovfieldmodel \approx \ones_{\nbsensors \times \nbsensors}$.
The condition \eqref{eq:condition_lowfreq} describes a low frequency regime, and is met when the array aperture is relatively smaller than the wavelength.

Since, to the best of our knowledge, no state-of-the-art work has addressed complex gain calibration from covariance readings, the vanilla solution will be used in the results \replaced{to benchmark the performance comparison}{as the baseline}. In the next sections, we focus on the proposed calibration method of this paper, which does not require a condition as strict as Eq. \eqref{eq:condition_lowfreq}.

\section{Complex gain calibration by covariance fitting}
\label{sec:sensor_calibration}

\subsection{The least square approach}
\label{subsec:ls_approach}

We propose to estimate $\vgains$ by solving a least-square problem. In the general case, the gain values are relative, \textit{i.e.} up to a complex constant. The least square minimization can be written by plugging in the sample covariance matrix in the sensing model \eqref{eq:sample_cov_total}:
\begin{equation}
\label{pb:leastsquare_vanilla}
\argmin_{\vgains} \dfrac{1}{2} \| \mcovsensesample - \diag(\vgains) \mcovfieldmodel \diag(\vgains)^\herm  \|^2_\frob ,
\end{equation}
the term $\varfield$ being discarded without loss of generality, since we seek relative gains in the first time -- see Sec. \ref{subsec:rel_to_abs} for more details. This form is non-convex over $\vgains$, and to our knowledge, no closed-form solution can be derived. Let $\lsloss$ be the least-square loss function; we remark that $\lsloss$ can also be written under the form
\begin{equation}
\label{eq:loss}
\lsloss(\mgains) = \dfrac{1}{2} \| \mcovsensesample - \mgains \odot \mcovfieldmodel \|^2_\frob,
\end{equation}
turning the input variable $\mgains=\vgains\vgains^\herm$ into a semi-definite positive matrix of rank 1. Thus, the least-square problem \eqref{pb:leastsquare_vanilla} can be recast in the following matrix form:
\begin{equation}
\label{pb:least_square_rank}
\mgainsest = \argmin_{\mgains \in \setpsd} \lsloss(\mgains) \text{ subject to } \rank(\mgains) = 1,
\end{equation}
with $\setpsd$ the set of semi-definite positive matrices. The relative gains $\vgainsest$ can be calculated straightforwardly from the principal eigenvector of $\mgainsest$. {\color{\revcolor}Consequently, since the latter is rank-one, a straightforward choice of $\vgainsest$ (still up to a constant) is \textit{e.g.} the first column of $\mgainsest$.} \deleted{By applying the eigenvalue decomposition [...]}

Note that problem \eqref{pb:least_square_rank} is still non-convex, now over $\mgains$, due to the rank constraint. Likewise, the loss in Eq. \eqref{eq:loss} can also be seen as a weighted case of matrix factorization with factors $\vgains$ and $\vgains^\herm$, and is known to be non-convex as well \cite{Davenport2016}. However, this reformulation is crucial for the sequel of the paper, because it unlocks the access to the low-rank matrix approximation, and the possibility to use efficient solvers accordingly.

\subsection{Penalized least-square and convex relaxation}
\label{subsec:convec_relaxation}

Finding the global minimizer of the problem \eqref{pb:least_square_rank} is generally a NP-hard problem \cite{Davenport2016}. However, it is possible to obtain a tractable alternative by applying convex relaxation to the program. To do so, the rank constraint on $\mgains$ is replaced by its convex envelope \cite{Fazel2001}: the nuclear norm $\|\mgains\|_\nucl$, defined by the sum of singular values of $\mgains$. This modification is enough to complete the convex relaxation, since (i) the set $\setpsd$ is already convex, and (ii) the loss term $\lsloss(\mgains)$ is convex on $\mgains$.

Rather than choosing a problem defined with constraint as in \eqref{pb:least_square_rank}, we favor a penalized least squares expression. We define the following problem \pbi as:
\begin{equation}
\tag{\pbi\!\!}
\label{pb:penalized_least_square}
\mgainsest = \argmin_{\mgains \in \setpsd} \lsloss(\mgains) + \regp \fpenalty_i(\mgains),
\end{equation}
where $\fpenalty_i(\mgains)$ is a penalty function (also known as a \textit{regularizer}) \added{defined for $i \in \{0, 1\}$}, and $\regp$ is a non-zero positive parameter that controls the importance of the penalty.

Let us define the convex form of the penalized least square for $i = 1$, \textit{i.e.} \pbone, associated to the penalty term $\fpenalty_1(\mgains) = \|\mgains\|_\nucl$. The global minimizer of \pbone can be easily found via a toolbox like CVX \cite{Grant2014}; however, the merit of using the penalized unconstrained form is to derive an efficient solver, such as proximal algorithm \cite{Parikh2014}, Alternating Direction Method of Multipliers \cite{Boyd2011}, or Frank-Wolfe algorithm \cite{Jaggi2013}. A solver for \pbone will be derived accordingly in section \ref{sec:solvers}.

Finally, for $i = 0$ the penalized least-square program \pbzero can be defined as an equivalent to Eq. \eqref{pb:least_square_rank}, if we choose  $\fpenalty_0(\mgains) = \indic(\mgains)$, such that $\indic(.)$ is the indicator function defined as \cite{Parikh2014}
\begin{align}
\begin{split}
\indic({\color{\revcolor}\bZ}) &= 0 \text{ if } {\color{\revcolor}\bZ} \in \setC, \\
            &= \infty \text{ otherwise,}
\end{split}
\label{eq:indicator_function}
\end{align}
and \replaced{$\setC = \{\bZ \in \setcomplex^{\nbsensors \times \nbsensors} | \rank(\bZ) = 1 \}$}{$\setC = \{\bM \in \setcomplex^{\nbsensors \times \nbsensors} | \rank(\bM) = 1 \}$}. Although with only a guarantee to find a local minimizer, we will see further in Sec. \ref{sec:solvers} that an efficient solver can be derived as well, because the problem \pbi is a generalized convenient form to switch between \pbzero and \pbone.

\subsection{From relative to absolute gains}
\label{subsec:rel_to_abs}

As seen in Sec. \ref{sec:vanilla} and \ref{subsec:ls_approach}, the provided estimation $\vgainsest$ is relative up to a constant in the general case. In some cases the relative solution could be enough (\textit{e.g.} in source localization). If one wants the absolute levels of sources, then a scaling correction is needed. We define $\vgainsestscaled = \eta \vgainsest$ the estimated gain vector after scaling, and $\eta$ the scaling correction. Below, we present some case examples of how to treat this issue.

\paragraph{Scaling with reference sensor(s)}

A first option is to consider the addition of one reference sensor. From asynchronous measurement of the power spectral density, it is then possible to estimate the variance of the diffuse field $\varfield$, because the latter is spatially homogeneous. In this case, we choose $\eta = \varfieldi$, yet the scaling is on amplitude only. Alternatively, an amplitude and phase correction is possible if the reference microphone is a part of the synchronous array (\textit{e.g.} if it is the 1-st microphone with gain 1, then $\eta = 1 / [\vgainsest]_1$). If several reference microphones are included, their gains can be incorporated as \textit{anchors} by a modification of the loss term -- see \cite[sec 3.2]{Balzano2008} for more details.

\paragraph{Scaling with respect to the gain value expectation}

A second option is to rely on the expected gain of the whole sensor collection. This is relevant when the manufacturer provides gain specifications with uncertainty (\textit{e.g.} the mean and standard deviation in the hardware datasheet). Let $\vgainsref$ be the vector of expected gains. The optimal correction in the least square sense aligns $\vgainsest$ onto $\vgainsref$ in the complex plane (like in Procrustes analysis \cite{Gower2010}). Then the scalar $\eta$ \replaced{is the minimizer of the least-square fitting problem}{minimizes}
\begin{equation}
    \label{pb:rel_to_abs}
    \eta = \argmin_{a \in \setcomplex} \|\vgainsref - a \vgainsest \|_2^2,
\end{equation}
so
\begin{equation}
    \label{eq:scaler_rel_to_abs}
    \eta = \dfrac{\vgainsest^\herm \vgainsref}{\|\vgainsest\|_2^2}.
\end{equation}
If all sensors are manufactured through the same process and have the gain mean value $\sgainmean$, then  $\vgainsref = \sgainmean \ones_\nbsensors$.

\paragraph{Scaling with respect to the ground truth}
\label{par:scaling_to_gt}

The last option is to use the ground truth as the reference, so $\vgainsref$ equals the true gain vector. This is irrelevant in practice, but is important to assess the performance of the calibration method only. Indeed, in that case, it is independent of a user-informed or \textit{anchor}-based scaling step. This will be used in result sections \ref{sec:numerical_experiment} and \ref{sec:experimental_results} to evaluate the performance of the calibration methods.

\section{Sensor gain calibration solvers}
\label{sec:solvers}

The problem \ref{pb:penalized_least_square} with the sum of a regression loss and a penalty term is a well-known general form met in many fields, such as machine learning \cite{Davenport2016} or computational imaging ill-posed problems \cite{Wei2020}. These applications usually meet large-scale problems, and require efficient solvers to deal with high dimensions. As a result, to tackle the non-smoothness of the penalty function, and to deal with large-scale problems, a plethora of dedicated optimization methods have been proposed. Below, we propose two solvers for \pbzero and \pbone, based on the first-order iterative proximal algorithm scheme \cite{Parikh2014,Beck2017}. \added{This approach is attractive because of the update rules are relatively simple, and the convergence rate is well-documented in the literature \cite{Beck2017}.}

The proximal gradient descent is an iterative algorithm based on the following update step \cite[ch. 10.2]{Beck2017}:
\begin{equation}
    \mgains^{[k+1]} = \prox_{\stepgrad \regp \fpenalty_i} \left(\mgains^{[k]} - \stepgrad \nabla \lsloss(\mgains^{[k]}) \right)
\end{equation}
where $k$ is the iteration index and $\stepgrad$ is the gradient step. It performs a gradient descent, followed by the proximal mapping with the operator $\prox_{\stepgrad \regp \fpenalty_i}(.)$ defined by the program
\begin{equation}
    \label{eq:prox_def}
    \begin{aligned}
        \prox_{{\color{\revcolor} \stepgrad \regp} \fpenalty_i} (\bZ) = \argmin_{\bM}  \frac{1}{2} \| \bM -\bZ \|^2_\frob + {\color{\revcolor} \stepgrad \regp} \fpenalty_i(\bM)
    \end{aligned}
\end{equation}
in the case of the mapping of a matrix $\bZ$, with $\regp$ being positive and $\gamma \nu$ resulting in some positive real scalar. As a result, the proximal algorithm is flexible, because solving \pbone and \pbzero only needs to choose the proximal operator accordingly. Finally, the loss gradient $\nabla \lsloss$ is obtained with the element-wise Wirtinger derivatives for $1 \leq m, n \leq \nbsensors$:
\begin{equation}
    \begin{aligned}
        \dfrac{\partial \lsloss(\mgains)}{\partial [\mgains]_{mn}^*}
        &= \dfrac{\partial}{\partial [\mgains]_{mn}^*}
        \left(
        \frac{1}{2}
        \left|
        [\mcovsensesample]_{mn} - [\mgains]_{mn} [\mcovfieldmodel]_{mn} \right|^2
        \right)
        \\
        &=
        - \frac{1}{2} [\mcovfieldmodel]_{mn}^*\left([\mcovsensesample]_{mn} - [\mgains]_{mn} [\mcovfieldmodel]_{mn} \right).
    \end{aligned}
\end{equation}
In the end, the gradient reads
\begin{equation}
    \nabla \lsloss(\mgains) = - \frac{1}{2} \mcovfieldmodel^* \odot (\mcovsensesample - \mcovfieldmodel \odot \mgains).
\end{equation}

\subsection{Resulting solver for \pbone}
\label{subsec:relaxed_pgd}

With the penalty term $\fpenalty_1(\bM)$, the proximal operator is given by
\begin{equation}
    \label{eq:prox_relaxed}
    \prox_{\stepgrad \regp \fpenalty_1} (\bZ) = \sum_{m=1}^{\nbsensors} (s_m - \stepgrad \regp)_+ \bu_m \bu_m^\herm
\end{equation}
where $(.)_+$ is the operator that clips the negative input values to $0$, and $s_m$ (resp. $\bu_m$) are the eigenvalues (resp. eigenvectors) of the matrix $\bZ \in \setpsd$. In the original form, the operator is defined by singular values and vectors, rather than the eigen-decomposition. However, the mapping is applied on $\mgains^{[k]} \in \setpsd$, and in this case singular value and eigenvalue decompositions are equal. The resulting solver is closely related to existing low-rank approximation methods for matrix completion -- see the singular value thresholding \cite{Cai2010}, or \cite{Ma2011}. The summary is described in the algorithm \ref{algo:prox} for $i = 1$.

The choice of parameter $\regp$ is crucial because it controls the rank sparsity of the estimation $\mgainsest$. First, it is possible to derive the upper bound value $\regp_{\text{max}}$ beyond which the rank is null, \textit{i.e.} $\mgainsest = \zeros_{\nbsensors \times \nbsensors}$. At convergence -- for $k \rightarrow \infty$ -- we have the equality $\mgains^{[\infty]} = \prox_{\stepgrad \regp_{\text{max}} \fpenalty_i} \left(\mgains^{[\infty]} - \stepgrad \nabla \lsloss(\mgains^{[\infty]}) \right)$. Solving the equation for $\mgains^{[\infty]} = \zeros_{\nbsensors \times \nbsensors}$ gives
\begin{equation}
\regp_{\text{max}} = \frac{1}{2} \|\mcovfieldmodel^* \odot \mcovsensesample\|_2.
\end{equation}
\deleted{where $\|\bM\|_2$ denotes the maximal singular value of $\bM$}. The authors suggest creating a regularization path, run the solver first with $\regp = \regp_{\text{max}} - 1\cdot10^{-6}$, then decrease the value of $\regp$ iteratively \added{(e.g. following a geometric series)}, and finally choose the smallest one which preserves $\mgainsest$ with rank one. By doing so, the biasing effect of the penalty term on the estimation is reduced. We emphasize that this suggestion is empirical, and that the optimal choice of $\regp$ still remains a general open problem.

\subsection{Resulting solver for \pbzero}
\label{subsec:nonconvex_pgd}

The penalty term given by the indicator function \eqref{eq:indicator_function} turns the proximal mapping in the Euclidean projection onto the set $\setC$ of $(\nbsensors \times \nbsensors)$ rank-1 matrices:
\begin{equation}
\label{eq:prox_proj_def}
\begin{aligned}
\prox_{{\color{\revcolor} \stepgrad \regp} \fpenalty_0} (\bZ)
&= \argmin_{\bM} \frac{1}{2} \| \bM -\bZ \|^2_\frob + {\color{\revcolor} \stepgrad \regp} \indic(\bM) \\
&= \argmin_{\bM \in \setC} \frac{1}{2} \| \bM -\bZ \|^2_\frob .
\end{aligned}
\end{equation}
If $\bZ$ has the \added{eigenvalue decomposition} (EVD) $\bZ =  \sum_{m=1}^{\nbsensors} s_m \bu_m \bu_m^\herm$, the Eckhart–Young–Mirsky theorem \cite[ch. 2.4]{Golub2013} proves that
\begin{equation}
\label{eq:prox_proj}
\prox_{{\color{\revcolor} \stepgrad \regp} \fpenalty_0} (\bZ) = s_1 \bu_1 \bu_1^\herm .
\end{equation}
Note that the operator is independent of $\regp$, because the indicator function takes either $0$ or $\infty$ values. For the notation thought, we keep it in the subscript in order to preserve the generalized notation from the definition \ref{eq:prox_def}. The resulting solver is in algorithm \ref{algo:prox}, with $i = 0$.

\subsection{Gradient step\replaced{,}{and} stopping criterion \added{and computational cost}}
\label{subsec:step_and_criterion}

The step $\stepgrad$ controls the speed of the gradient descent. The stable convergence is determined by the interval $\stepgrad \in ]0, \frac{1}{\mathcal{L}}]$, where $\mathcal{L}$ is the Lipschitz constant of the gradient $\nabla \lsloss$. Here it can be easily shown that $\displaystyle \mathcal{L} = \frac{1}{2} \max_{m, n}(|[\mcovfieldmodel]_{mn}|^2)$. If the goal is to maximize the convergence speed without falling into instability, $\stepgrad$ should be the highest one in this interval. With the diffuse field model \eqref{eq:model_cov} it boils down to $\stepgrad = 2$.

For the convergence of the algorithm, we rely on the evolution of $\mgains^{[k]}$ and arbitrarily choose $\|\mgains^{[k]} - \mgains^{[k-1]}\|_\frob \leq 10^{-6}$ \added{as a stopping criterion}.  Note that the convergence rate of low-rank approximation with non-convex penalties remains an open problem (here the solver for \pbzero\!\!); but interestingly, empirical studies evidence their faster convergence than when using convex ones \cite{Yao2019}. \added{In terms of computational cost, the solvers for $i=0$ and $i=1$ are mostly explained by the convergence rate because their update rule have the same complexity. However when $i=1$, the regularization parameter is generally unknown and the a regularization path must be performed by running the algorithm for different values of $\nu$. In the latter case, the computation cost is higher.}

\begin{algorithm}[h]
\caption{sensor calibration in ambient noise by proximal gradient descent}
\label{algo:prox}
\SetKwInOut{Input}{input}
\SetKwInOut{Output}{output}
\SetKwFunction{Evd}{evd}
\Input{$\vgains^{[0]}$, $\stepgrad$, $\regp$, $i$, $k_{\text{max}}$}
initialize $\mgains^{[0]} = \vgains^{[0]} {\vgains^{[0]}}^{\herm}$, $k=0$\;
\Repeat{$\|\mgains^{[k+1]} - \mgains^{[k]}\|^2_\frob < 10^{-6} \mathrm{~or~} k = k_{\mathrm{max}}$}
{$\mgains^{[k+1]} = \prox_{\stepgrad \regp \fpenalty_i} \left(\mgains^{[k]} - \stepgrad \nabla \lsloss(\mgains^{[k]}) \right)$\;
$k \leftarrow  k + 1$\;}
obtain the EVD  $\mgains^{[k]} = \sum_{k=1}^{\nbsensors} s_i \bu_i \bu_i^\herm$\, and take $\vgainsest = \bu_1$\;
[optional: compute $\eta$ for relative to absolute scaling]\;
\Output{$\vgainsest$ [optional: $\vgainsestscaled = \eta \vgainsest$]}
\end{algorithm}

\section{Numerical analysis of the performance}
\label{sec:numerical_experiment}

The aim of this section is to first study the performance and the robustness of the proposed gain estimators. Concerning performance, we analyze how the signal-to-noise ratio (SNR) and the number of snapshots in the sample covariance matrix affect the calibration results. With linear array geometries, we will also study how the performance depends on the wavelength $\wavelength$. Concerning robustness, we will evaluate the calibration sensitivity to mismatch with the exact diffuse noise model. Three methods will be compared: the proximal gradient descent algorithm \ref{algo:prox} for \pbone and \pbzero (referred to as pgd-\pbone and pgd-\pbzero respectively), and the vanilla estimator from Sec. \ref{sec:vanilla} (referred to as vanilla).

\added{Performed in the frequency domain,} the simulation generates snapshot measurements $\vsense_l$, and the field vectors $\vfield_l$ result in the propagation of uncorrelated plane waves uniformly distributed on the sphere. The SNR of snapshots respects the ratio
\begin{equation}
    \label{eq:snr}
    \text{SNR} = 10 \log_{10} \left(\dfrac{\expectancy \{\| \diag(\vgains) \vsense_l \|^2\} }{ \expectancy \{ \|\vnoise_l\|^2\} } \right),
\end{equation}
the ambient noise variance is fixed at $\varfield = 1$, and the sensor noise variance $\varnoise$ is tuned to match the SNR according to \eqref{eq:snr}. The gains are scaled with respect to the ground truth (see Sec. \ref{subsec:rel_to_abs}, paragraph c). Results are provided either by the root-mean-square error  $\text{RMSE} = \|\vgainsref - \vgainsestscaled\|_2$, or by the relative RMSE: $\text{rRMSE} = \|\vgainsref - \vgainsestscaled \|_2 / \|\vgainsref\|_2$. In practice, the rRMSE is interesting because it is simple to convert a relative error to a measured level accuracy or a frequency response uncertainty in dB, as it is usually provided in a microphone datasheet. For example, a $10 \%$ relative error corresponds to $\pm 0.9$ dB. \added{More importantly, the reader can convert errors to dB and compare with the typical uncertainties of standard calibration methods for microphones \cite{Frederiksen2013}.}

Note that in the current particular case where $\vgainsref$ is the ground truth, it is easy to show that
\begin{equation}
\text{rRMSE} = \sqrt{1 - \rho^2}
\end{equation}
where $\rho = |\vgainsref^\herm \vgainsest| / (\|\vgainsref\| \|\vgainsest\|)$ is the correlation coefficient between the estimated gains and the ground truth. So here the rRMSE is bounded between $0$ and $1$, and the worst (resp. best) estimator is when rRMSE = $1$  (resp. rRMSE = $0$) with high probability, i.e. with a little standard deviation of this rRMSE.

\subsection{Effect of SNR and snapshot number}
\label{subsec:numexp_snapsnr}

In each simulation, the array is planar, forms a $\SI{1}{\meter}$ diameter disk in which $\nbsensors = 50$ sensors are randomly positioned. The complex gains $[\vgains]_m$ are one-mean, the real and imaginary parts uniformly distributed in the intervals $[0.5, 1.5]$ and $[-0.5, 0.5]$ respectively. We consider an acoustic scenario with $\soundspeed = \SI{343}{\meter\per\second}$ at frequency $\SI{500}{\hertz}$, so $\wavelength = \SI{68.6}{\centi\meter}$. Statistical results provide the mean and standard deviations of rRMSEs from $500$ simulations.

\begin{figure}
\figcolumn{
\fig{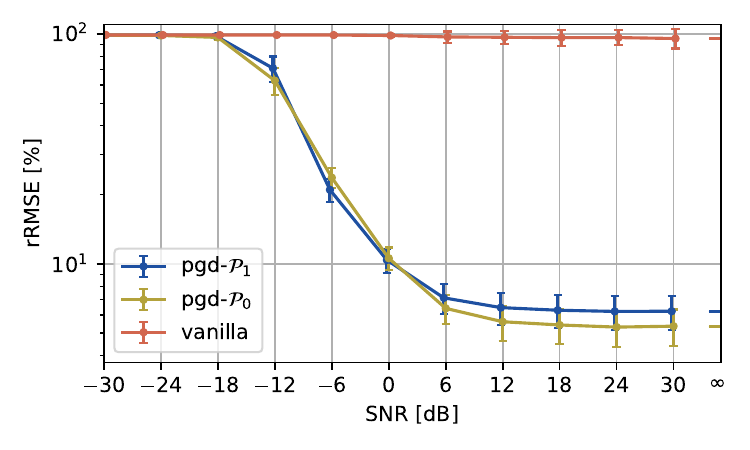}{\reprintcolumnwidth}{(a)}{\label{fig:simu_snr}}
\fig{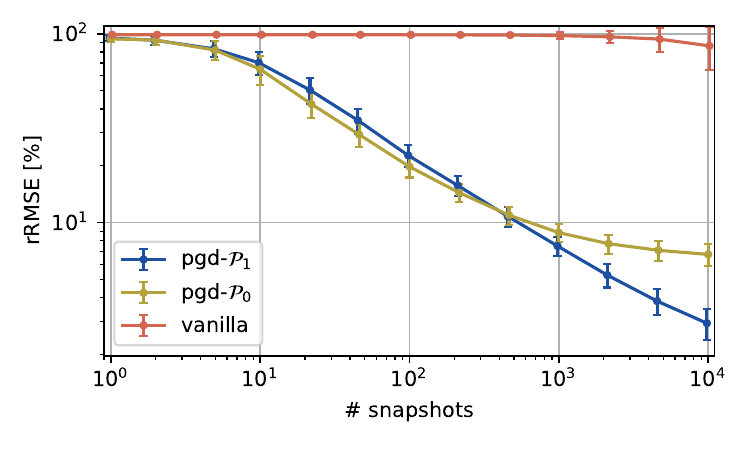}{\reprintcolumnwidth}{(b)}{\label{fig:simu_snap}}
}
\caption{(color online) Complex gain calibration of $M= 50$ sensors in ambient noise. rRMSE means and standard deviations from 500 simulations, with (a) $L = 500$ snapshots, (b) SNR $= 0$ dB. Solvers: \pbone (blue), \pbzero (green) and vanilla (red).}
\label{fig:simu_snrsnap}
\end{figure}

The parametric study on SNR \replaced{is}{if} given in Fig. \ref{fig:simu_snr}, for $\nbsnap = 500$ snapshots. The SNR ranges from $-30$ to $\SI{30}{\decibel}$ and includes the asymptotic case $\nbsnap = \infty$ \replaced{(see Fig. \ref{fig:simu_snr} with}{-- cf.} horizontal bars on the right\added{)}. The trends for pgd-\pbone and pgd-\pbzero  are close, with a slightly better rRMSE for pgd-\pbzero in high SNR. The vanilla approach provides irrelevant estimations for all SNRs in the given configuration.
For the parametric study on snapshots plotted in Fig. \ref{fig:simu_snap}, $\nbsnap$ spans from 1 to 10000, and $\text{SNR} = \SI{0}{\decibel}$. Again, the vanilla estimation fails, with a slight decrease for large $\nbsnap$. Methods pgd-\pbone and pgd-\pbzero also keep close trends, but with a large number of snapshots pgd-\pbone keeps decreasing the rRMSE and performs better.

The two figures reveal the potential instability of the vanilla estimation, as discussed in Sec. \ref{sec:vanilla}. The proposed algorithm is stable (\textit{i.e.} with little standard deviation). Despite the difficulty of solving a non-convex problem, pgd-\pbzero shows surprisingly good results, although the two parametric studies evidence marginal differences between pgd-\pbzero and pgd-\pbone except in extreme regimes -- for SNR or $\nbsnap$ large.

\subsection{The importance of spatial over-sampling}
\label{subsec:numexp_wavelength}

The diffuse noise model expression evidences that the array geometry can affect the least-square loss. Indeed, the term $\mcovfieldmodel \odot \mgains$ weights the unknown matrix by a $\sinc$ function, taking zero values (when $\dist_{mn} = k \wavelength / 2 $ for $k \geq 1$) and quickly decaying as $\dist_{mn} / \wavelength$ grows.

\begin{figure}
    \figcolumn{
        \fig{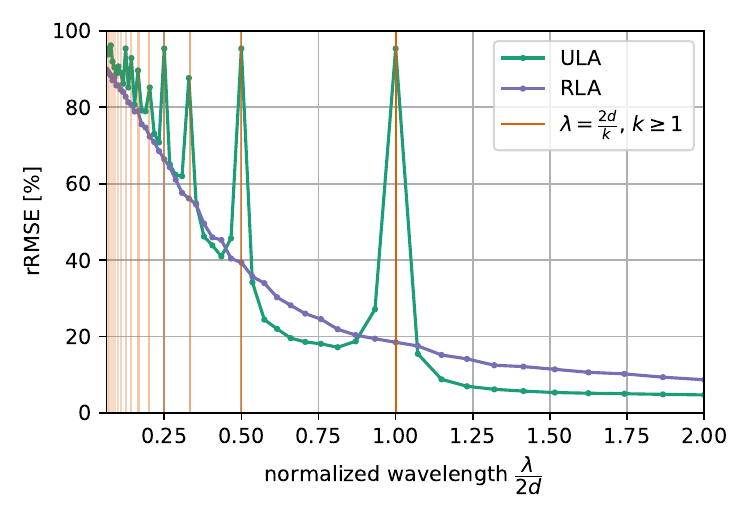}{\reprintcolumnwidth}{(a)}{\label{fig:simu_wavelength_MC}}
        \fig{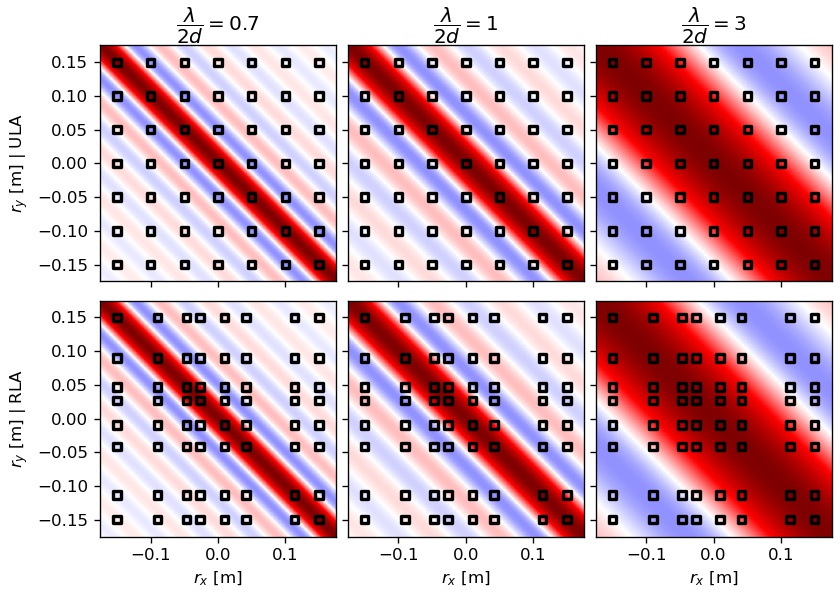}{\reprintcolumnwidth}{(b)}{\label{fig:simu_wavelength_explained}}
    }
    \caption{(color online) Analysis of the diffuse field spatial sampling. (a) Complex gain calibration of uniform and random linear arrays with $M=21$, aperture $\SI{1}{\meter}$. ULA inter-element spacing $d = \SI[]{5}{\centi\meter}$. Mean rRMSE on 500 simulations for each point, with pgd-\pbone\!\!. (b) Illustration in the 1-dimensional case of the 3 regimes: over-sampling (right), critical sampling (middle), and sub-sampling (left) of the ambient field covariance with a 8-element ULA (top) and RLA (bottom).}
    \label{fig:simu_wavelength}
\end{figure}

For illustration purpose only, Fig. \ref{fig:simu_wavelength_explained} depicts a 1-dimensional scenario with a uniform linear array (ULA) and a random linear array (RLA). It draws the continuous cross-spectral density on the line between two points of coordinate $r_x$ and $r_y$, and has the amplitude $\sinc(2 \pi |r_x - r_y| / \wavelength)$. The black squares represent the spatial sampling obtained by a 8-element ULA (top) and RLA (bottom), so that the resulting $8 \times 8$ matrix $\mcovfieldmodel$ consists of the values indicated in these squares. The uniform sampling (top row) clearly reveals 3 distinct regimes,  depending on whether the spatial Shannon-Nyquist criterion is met or not \cite{Menon2012}. Let $\dist$ be the ULA inter-element spacing, then over-sampling (resp. sub-sampling) is met when $d / \wavelength < 1 / 2$ (resp.  $d / \wavelength > 1 / 2$). Critical sampling occurs when the off-diagonal black squares are located to zeros. In this case, we have $\mcovfieldmodel = \mId_\nbsensors$. This situation also occurs when $\wavelength = 2 d / k$ for $k \geq 1$.

Figures \ref{fig:simu_wavelength_MC} shows the consequence of spatial sampling as a function of wavelength, with a $21$-element ULA and RLA of aperture $\SI{1}{\meter}$  (aperture is defined by $\max_{m, n}(\dist_{mn})$), SNR = $\SI{15}{\decibel}$, and $\nbsnap = 1000$. The averaged rRMSE with the solver for \pbone are plotted as a function of the normalized wavelength. When $\mcovfieldmodel = \mId_\nbsensors$ (indicated with orange vertical lines) the estimation fails for the ULA. Indeed, the loss as initially defined in Eq. \eqref{eq:loss} is reduced to
\begin{equation}
\begin{split}
\lsloss(\vgains\vgains^\herm)
&= \frac{1}{2} \| \mcovsensesample - \mId_\nbsensors \odot \vgains\vgains^\herm \|^2_\frob \\
&= \frac{1}{2} \sum_{m=1}^{\nbsensors} \left( [\mcovsensesample]_{mm} - |[\vgains]_m|^2 \right)^2 ,
\end{split}
\end{equation}
and since covariance are in $\setpsd$ the diagonal terms are real-valued, therefore the estimation suffers from phase ambiguities on the gains. The problem does not arise with RLA, because $\mcovfieldmodel$ is never diagonal -- see Fig. \ref{fig:simu_wavelength_explained}. However, for both ULA and RLA the performance decays in sub-sampling because $[\mcovfieldmodel]_{mn}$ ($m \neq n$) fades to $0$ as $\lambda$ decreases. As a result, the ideal configuration to obtain low rRMSE remains in this so-called over-sampling regime for any array geometry, which arises when the main lobe of the spatial sinc function contains more than one sampling point. \added{Note that this limitation also corroborates with the required condition for the Vanilla method in the Eq. \ref{eq:condition_lowfreq}.}

\subsection{Model mismatch: a case study}
\label{subsec:numexp_modelmismatch}

\begin{figure}
    \figcolumn{
        \fig{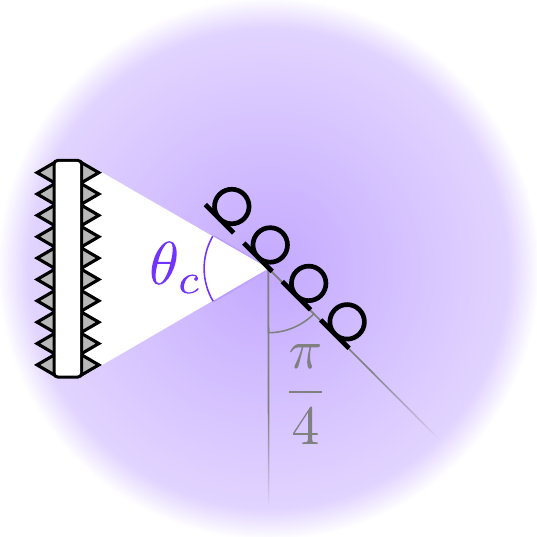}{0.4\reprintcolumnwidth}{(a)}{\label{fig:simu_modelmismatch_scheme}}
        \fig{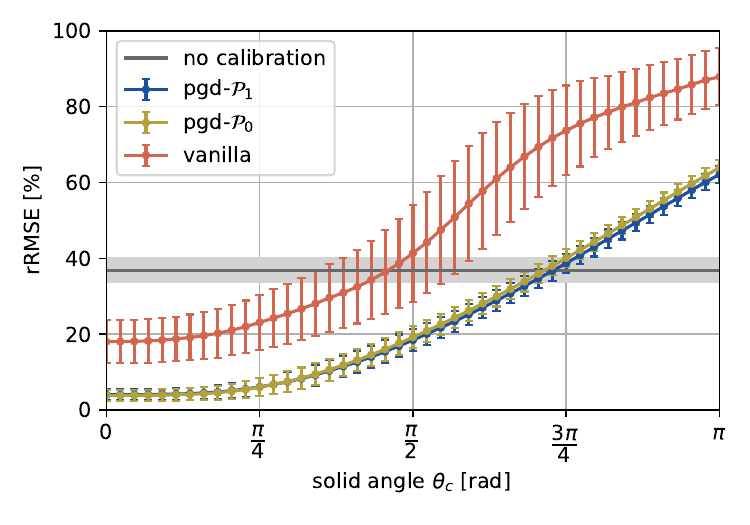}{\reprintcolumnwidth}{(b)}{\label{fig:simu_modelmismatch_mc}}
    }
    \caption{(color online) Sensor calibration in presence of model mismatch: (a) retina absorbs plane waves in a solid angle $\theta_c$. Illustration with $\nbsensors=4$. (b) rRMSE means and standard deviations on 500 simulations, with SNR $= \infty$ and $\nbsnap = 500$.}
    \label{fig:simu_modelmismatch}
\end{figure}

We want to evaluate the sensitivity of the calibration due to the model mismatch from the diffuse field property. To do so, we perturb the uniform spatial distribution of plane waves by adding an anechoic retina in the far field, see the geometric configuration in Fig. \ref{fig:simu_modelmismatch_scheme} (4 sensors are drawn only for the visibility of the illustration). The size of the retina is adjusted by the solid angle $\theta_c \in [0, \pi]$, the case $\theta_c = \pi$ depicting the scenario where the array is nearby an anechoic wall. Such a partially diffuse field was studied analytically in \cite{Blake1977}, but we consider this mismatch as unknown here.

Figure \ref{fig:simu_modelmismatch_mc} compares the rRMSE means and standard deviations for the 3 solvers, plus the case in which no calibration is performed in gray. There is no sensor noise (\textit{i.e.} $\varnoise=0$), and we keep the ULA configuration from the previous section and the wavelength is reset to $\wavelength = \SI{68.6}{\centi\meter} $ -- in the spatial over-sampling regime. As depicted in  Fig. \ref{fig:simu_modelmismatch_scheme}, the retina is oriented $ \SI[parse-numbers=false]{\pi/4}{\radian} $ from the ULA principal axis \deleted{to discard symmetries}. In this configuration, the vanilla approach works better than in Fig. \ref{fig:simu_snrsnap} but remains more sensitive to model mismatch, with a more significant rRMSE increase than with proximal solvers. As expected, it remains less stable, with larger standard deviations. The solver pgd-\pbone is marginally better than pgd-\pbzero when $\theta_c$ is large, and both do not suffer from the mismatch when $\theta_c \leq \pi/4$. The comparison with uncalibrated gains (gray) gives an indication of the range of angles $\theta_c$ where the calibration brings an improvement or not; nevertheless it should not be taken as a generality because the rRMSE level in gray is only representative of the gain distribution that was chosen -- cf. Sec. \ref{subsec:numexp_snapsnr}.

\section{Experimental results}
\label{sec:experimental_results}

\begin{figure}
\centering
\includegraphics[width=0.7\reprintcolumnwidth]{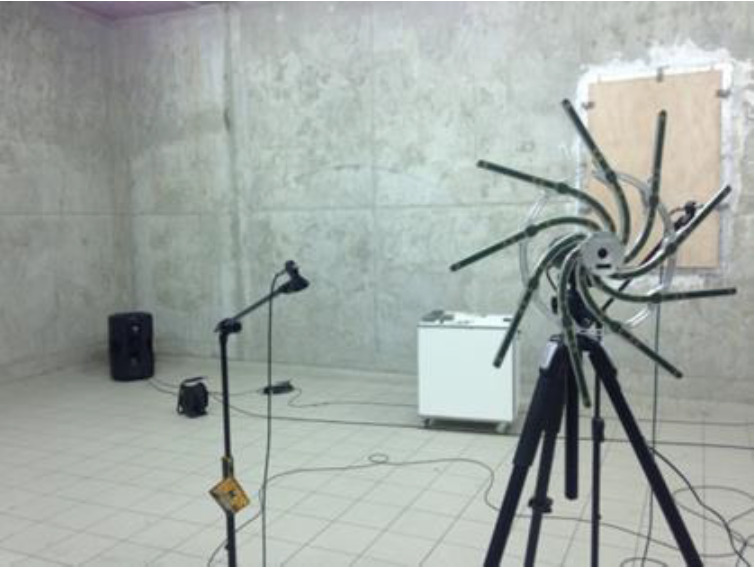}
\caption{(color online) Experimental setup: a 43-microphone array in a reverberant room.}
\label{fig:exp_setup}
\end{figure}

This second section of results evaluates the ability of gain calibration on acoustic real data captured by a ``Simcenter Sound Camera" antenna from Siemens. The scenario is the following: a 43-microphone spiral array is \SI{30}{\centi\meter} wide, and records acoustic pressures in a reverberant room at the sampling frequency of \SI{51.2}{\kilo\hertz}. We emphasize that the microphone array is already calibrated off-the-shelf by Siemens, so we deliberately assign gain coefficients on the measurements in the presented results below. The picture of the experiment is provided in Fig. \ref{fig:exp_setup}. First the quality of the diffuse field is investigated from the raw measurements, then the calibration step is performed and assessed.

\subsection{Ambient noise conditions}
\label{subsec:exp_raw}

\begin{figure*}
    \fig{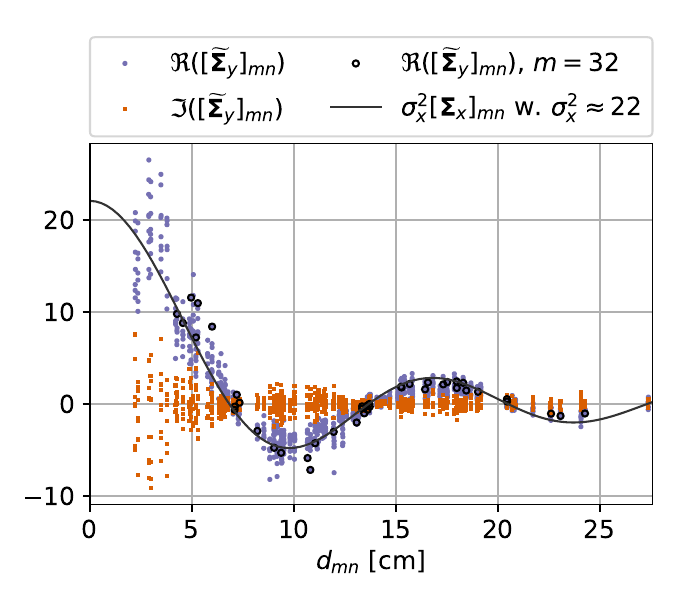}{0.35\textwidth}{(a)}{\label{fig:exp_csmillustratedA}}
    \fig{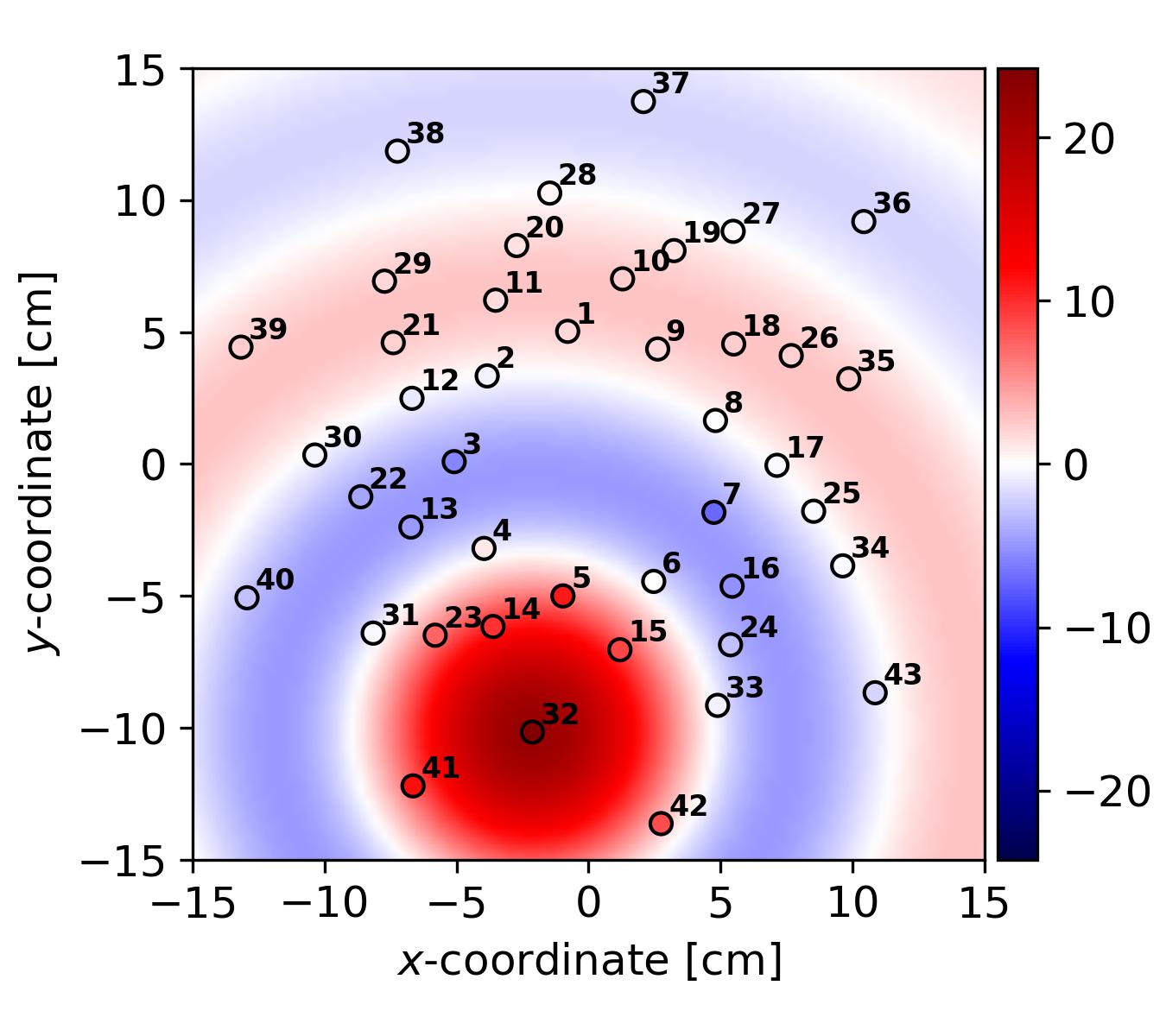}{0.35\textwidth}{(b)}{\label{fig:exp_csmillustratedB}} \\
    \fig{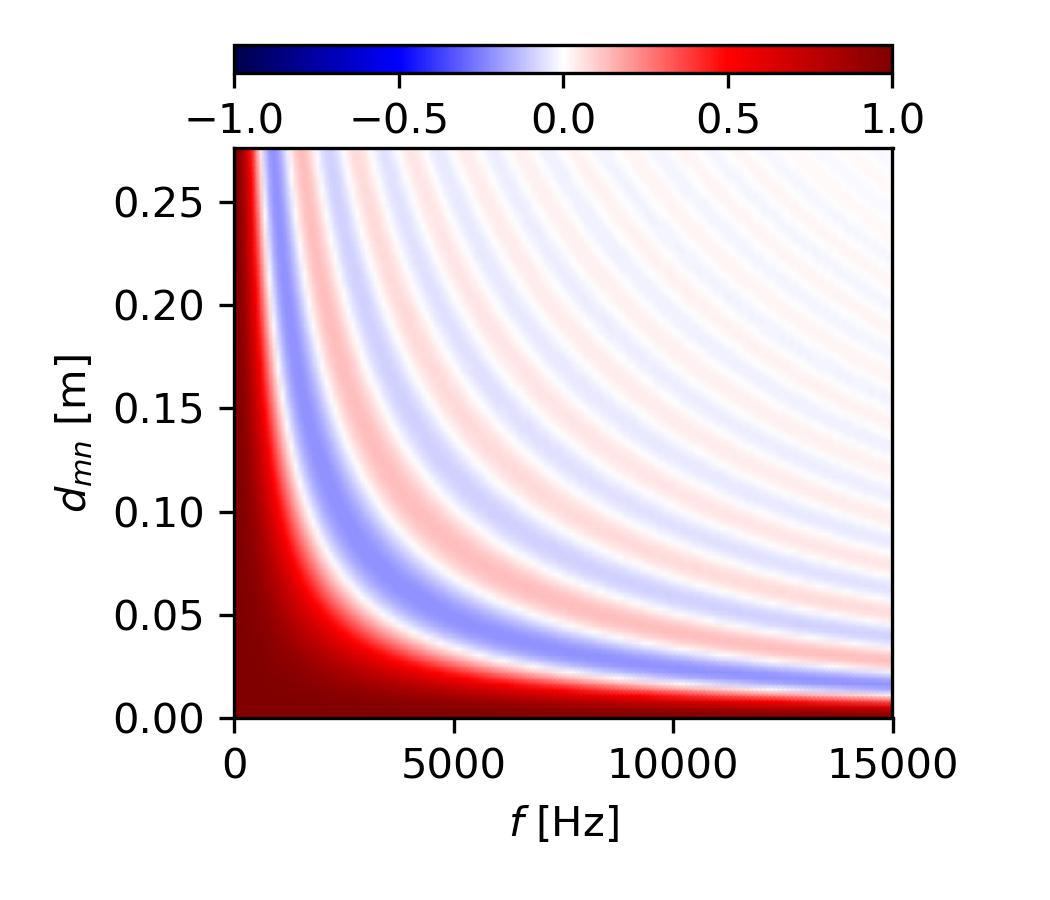}{0.35\textwidth}{(c)}{\label{fig:exp_csmillustratedC}} ~
    \fig{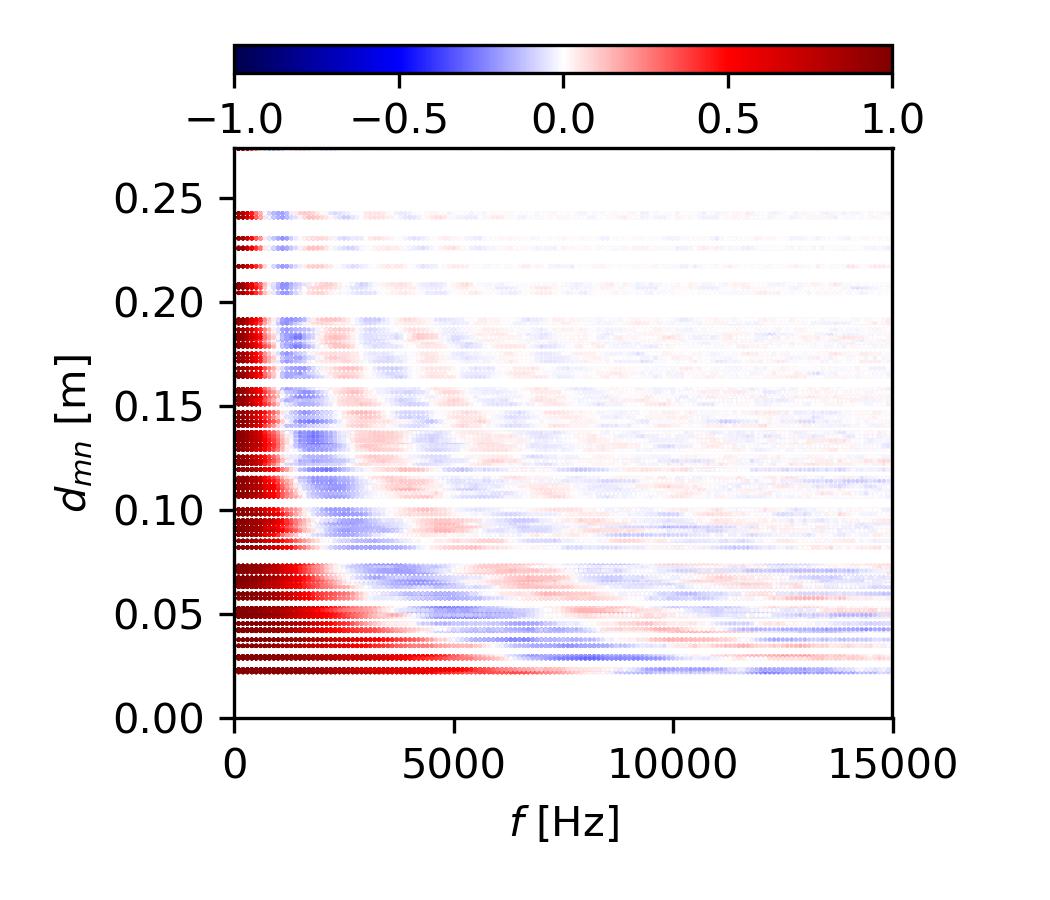}{0.35\textwidth}{(d)}{\label{fig:exp_csmillustratedD}}
    \caption{(color online) Sample covariance $\mcovsensesample$ off-diagonal coefficients before calibration in the reverberant room (a) $\mcovsensesample$ off-diagonal matrix coefficients at $f= \SI{2.5}{k\hertz}$; black circles indicate values paired with the $32$-nd microphone. (b) Spatial covariance with respect to the $32$-nd sensor; analytical value (background) and sampled values (in circles). Real part of (c) the analytical scaled covariance with $\dist_{mn} \in \SI[parse-numbers = false]{[0, 0.27]}{\meter}$ , and of (d) the $\mcohsensesample$ coefficients, in the (frequency, distance) plane.},
    \label{fig:exp_csmillustrated}
\end{figure*}

The field is known to be diffuse in reverberant rooms above the \textit{Schroeder frequency} \cite{Pierce2019,Nelisse1997}. Here, the room has a volume of \SI[]{411}{\cubic\meter}, a reverberation time of about \SI{9}{\second}, and a Schroeder frequency of about \SI{0.3}{k\hertz}. For frequency to wavelength conversion, we measured the sound speed $\soundspeed = \SI{341.5}{\meter\per\second}$. The diffuse sound field is excited by two speakers positioned at room corners and emitting white noise. From a \SI{42}{\second}-long recording, the sample covariance matrix $\mcovsensesample$ is computed with $\nbsnap = 6008$ snapshots. To do so, the DFT is applied to time series blocks of $512$ samples with frequency resolution $\Delta f = \SI{100}{\hertz}$, thereby the condition \eqref{eq:condition_dft} is respected.

The raw measurements before calibration are visualized in Fig. \ref{fig:exp_csmillustrated}. First, Fig. \ref{fig:exp_csmillustratedA} shows the real and imaginary parts of $[\mcovsensesample]_{mn}$ coefficients at $f= \SI{2.5}{k\hertz}$. Since $\mcovsensesample$ is hermitian, and since the diagonal elements involve variances, the coefficients from the upper triangular part are plotted only, that is for $n > m$. For comparison with the model, the analytical expression is plotted in green, with an approximation of $\varfield$ based on the average of the measured variances (located in the diagonal of $\mcovsensesample$). We recall that the analytical imaginary part is null. The measured covariance points are well concentrated and close to the model. The dispersion of these points around their average trend are either due to the dispersion of the sensor gains, or to the mismatch between the model and the sensed sound field.

\begin{figure}
    \includegraphics[width=\reprintcolumnwidth]{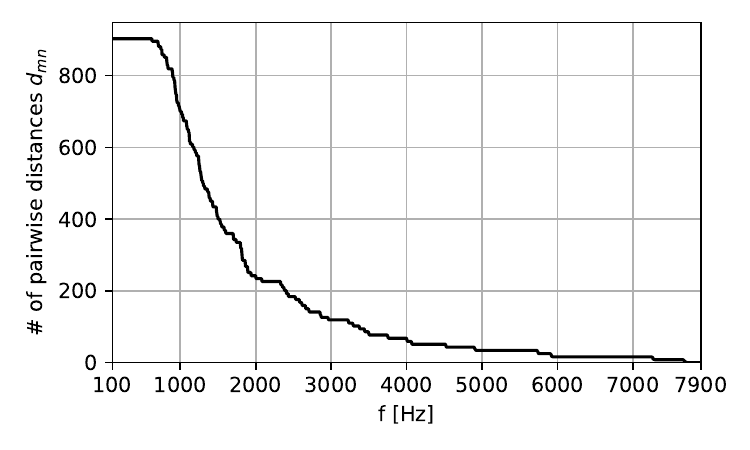}
    \caption{(color online) Evaluating the over-sampling with the experimental microphone array: number of pairwise distances $\dist_{mn}$ such that $\dist_{mn} / \wavelength < 1/2$, $\wavelength = \soundspeed / f$.}
    \label{fig:exp_oversampling}
\end{figure}

In the 2-dimensional case with irregular geometry, determining over-sampling regime described in Sec. \ref{subsec:numexp_wavelength} is not trivial. Yet,
\begin{itemize}
    \item a first clue to help answer this question is given in Fig. \ref{fig:exp_csmillustratedB}, still for $f= \SI{2.5}{k\hertz}$. The background plots the analytical covariance between the $32$-nd microphone position and a second coordinate within a $\SI{30}{\centi\meter} \times \SI{30}{\centi\meter}$ square. The circled points are filled with the real part measured values, that is $\Re([\mcovsensesample]_{mn})$ for $m = 32$ and $n \neq m$. Note that these values correspond to the ones circled in black in Fig. \ref{fig:exp_csmillustratedA}. The over-sampling was identified with ULAs when $\wavelength / (2 d)$. The visual interpretation of this in Fig. \ref{fig:exp_csmillustratedB} is that nearby sensors are present in the main lobe of the sinc-shaped spatial covariance. This is true when the covariance is centered at the $32$-nd sensor, and seems to hold if centered at other microphone positions. Nevertheless, this visual interpretation is valid for $f= \SI{2.5}{k\hertz}$ only;
    \item to generalize over other frequencies, a second clue is provided in Fig. \ref{fig:exp_oversampling}. It draws, as a function of $f = \soundspeed / \wavelength$, the number of pairwise distances respecting the condition $\dist_{mn}/\wavelength > 1/2$. This number quickly decays from $f = \SI{0.8}{k\hertz}$ to \SI{3}{k\hertz}, and reaches 0 at \SI{7.7}{k\hertz}. As a rule of thumb, at least \SI{10}{\percent} of the distances \replaced{should be}{respect} below $f = \SI{3.4}{k\hertz}$.
\end{itemize}

Finally, Figs. \ref{fig:exp_csmillustratedC} and \ref{fig:exp_csmillustratedD} provide a complete side-by-side comparison of the model with all the measurements in the $(f, \dist_{mn})$ plane. Since $\nbsensors = 42$ there are $\nbsensors(\nbsensors - 1)/2 = 861$ pairwise distances; with $79$ frequency bins from $0.1$ to \SI{7.9}{k\hertz}, Fig. \ref{fig:exp_csmillustratedD} consists of $68019$ points overall. Rather than directly relying on the covariance, we plot the real part of the \replaced{correlation}{coherence}. By definition, the \replaced{correlation}{sample coherence} matrix $\mcohsensesample$ is a normalization of the sample covariance coefficients $[\mcovsensesample]_{mn}$, that is
\begin{equation}
[\mcohsensesample]_{mn} = \dfrac{[\mcovsensesample]_{mn}}{\sqrt{[\mcovsensesample]_{mm} [\mcovsensesample]_{nn}}}.
\end{equation}
Although the speakers generates white noise their frequency response is not flat, thus the source variance $\varfield$ varies across frequencies. Microphone responses are not flat as well. This is why plotting the \replaced{correlation}{coherence} is more convenient: it reveals the \added{actual} ambient noise spatial correlation by mitigating the hardware frequency responses and the variations of $\varfield$ over frequencies. Note that microphone amplitude part of the gains are discarded, but not the phase part. The side-by-side comparison of the real part of analytical and measured \replaced{correlations}{coherences} reveal a good similarity in the considered frequency range.

In summary, this section validates that the diffuse noise model is a reliable choice to perform the proposed sensor gain calibration, in the acoustic scenario with a reverberant room. Although the array geometry is irregular and planar, a proxy was proposed to identify an approximate frequency range where spatial over-sampling occurs, that is below \SI{3.5}{\kilo\hertz}.

\subsection{Gain calibration results}
\label{subsec:exp_results}

\begin{figure}
    \figcolumn{
        \fig{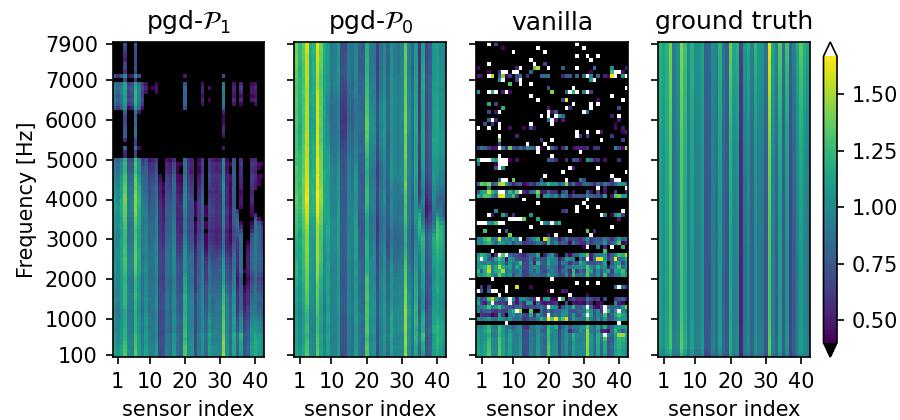}{\reprintcolumnwidth}{(a)}{\label{fig:exp_mapgains_coeff}}
        \fig{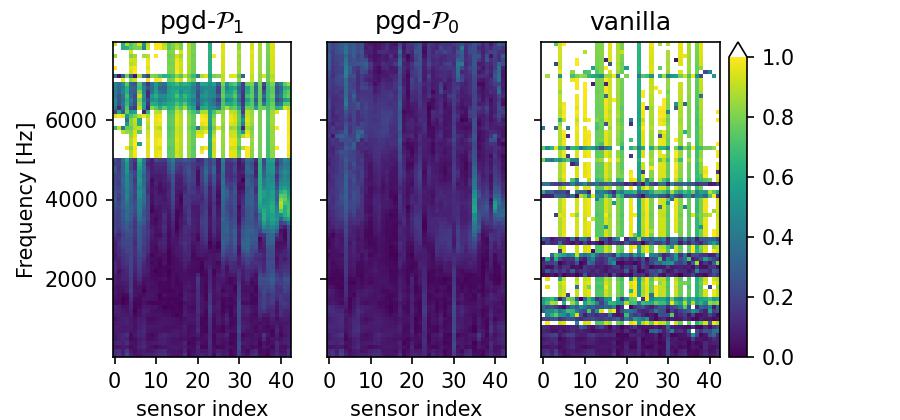}{\reprintcolumnwidth}{(b)}{\label{fig:exp_mapgains_relerr}}
    }
    \caption{(color online) Experimental sensor gain calibration: (a) amplitude gains $|[\vgainsestscaled]_m|$, (b) relative error for each sensor and frequency.}
    \label{fig:exp_mapgains}
\end{figure}

In this section, we perform the microphone gain calibration from $0.1$ to \SI{7.9}{k\hertz}. Since the sensing model is expressed for a given frequency, the calibration is launched for each frequency bin, resulting in $79$ vectors $\vgainsest$, or $79 \times 43$ coefficients overall. The scaling is done by choosing $\vgains_0$ as the ground truth, which is known since gains were artificially applied on the snapshots readings. {\color{\revcolor} In order to determine the regularization parameter for pgd-\pbone, a regularization path is obtained by decreasing $\regp$ with a geometric series of common ratio 0.98, and the collected rank-one solution $\mgainsest$ with the smallest $\regp$ is kept}.

The absolute value of these coefficients are drawn in Fig. \ref{fig:exp_mapgains_coeff}, ground truth $\vgains_0$ included. Additionally, Fig. \ref{fig:exp_mapgains_relerr} shows the coefficient-wise relative errors with respect to the ground truth $|[\vgainsestscaled]_m - [\vgains_0]_m| / |[\vgains_0]_m|$. First, the vanilla approach is the least reliable one, because most estimated gains are irrelevant, except in low frequencies and around \SI{2.5}{k\hertz}. Concerning low frequencies, the estimations are correct because the condition \eqref{eq:condition_lowfreq} is respected (cf. Fig. \ref{fig:exp_oversampling}) and stabilizes the element-wise division by $\mcovfieldmodel \approx \ones_{\nbsensors \times \nbsensors}$. However, the proposed algorithm performs better on a wider range of frequencies:
\begin{itemize}
    \item with pgd-\pbone the estimation works until \SI{5}{\kilo\hertz}; and fails beyond. Some errors are visible from $3$ to \SI{5}{\kilo\hertz} on the peripheral microphones, associated to large sensor indexes;
    \item the estimations by pgd-\pbzero are good on the whole studied frequency range, with a homogeneous relative error per sensor and per frequency. A local increase remains visible in the sector where $f \in \SI[parse-numbers=false]{[3, 5]}{\kilo\hertz}$ and where sensor index $m \geq 36$.
\end{itemize}

\begin{figure}
\includegraphics[width=\reprintcolumnwidth]{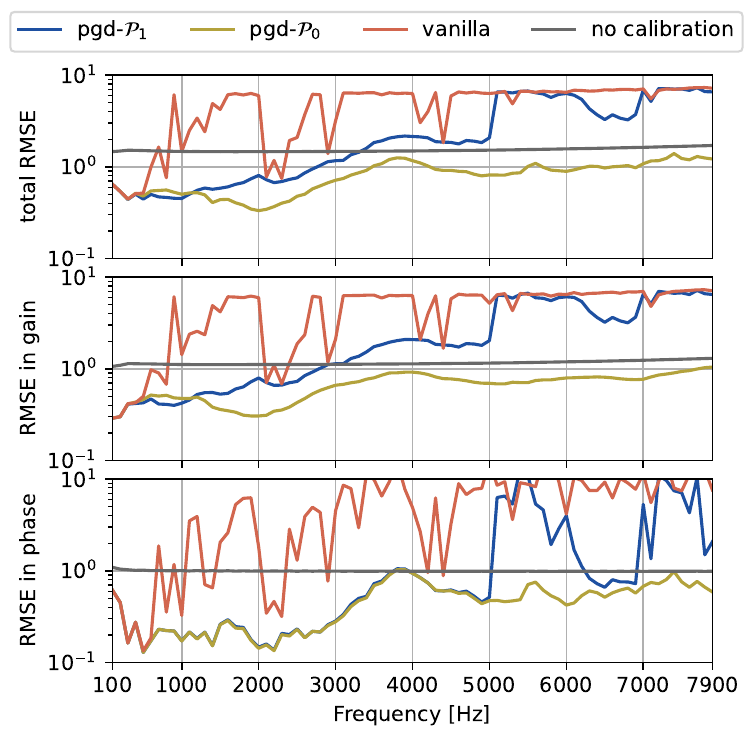}
\caption{(color online) Experimental microphone gain calibration RMSE: total (top), amplitude only (middle), and phase only (bottom).}
\label{fig:exp_rmsefreqall}
\end{figure}

Finally, the performance per frequency is given in Fig. \ref{fig:exp_rmsefreqall}. In order to discriminate the nature of errors, the RMSE on raw complex gains (top), their absolute values (middle) and their phase (bottom) are plotted. The gray graphs indicate the RMSE if no calibration was performed. The RMSE are the same for the three compared methods below \SI{600}{\hertz}, and quickly diverge beyond. Interestingly, the error in phase are identical for pgd-\pbzero and pgd-\pbone, until  pgd-\pbone fails from \SI{5}{\kilo\hertz}. Thus, the difference between both is mainly explained by the amplitude component. Again, pgd-\pbzero reveals to be the most stable: the total RMSE is small in low frequencies, slightly increases from $2$ to \SI{3.8}{\kilo\hertz}, and remains stable beyond. Note that this trend is more obvious on the phase RMSE. The authors suggests explaining this trend by the spatial sampling density, as it was discussed previously. Indeed, the proxy indicator Fig. \ref{fig:exp_oversampling} reveals that over-sampling is marginal beyond \SI{3.5}{\kilo\hertz}. This frequency interval matches the one in which RMSE is lower. Thus, this experimental analysis confirms again that the relative dimension between the array geometry and the wavelength drives the calibration performance, even with a relevant model choice.

\section{Conclusion}

The complex gain calibration of sensors from a synchronous array in ambient noise is possible by leveraging the spatial structure of the cross-spectral measurements. Unlike the conventional choice of snapshot readings, it relies on second-order statistical measurements: the problem is cast in the least-square form to fit the diffuse noise model with the sample covariance matrix. The problem is non-convex, and can be modified by convex relaxation via the low-rank matrix approximation framework.

A proximal algorithm was proposed, in order to deal efficiently with $\nbsensors$ large, and to solve the problem with or without convex relaxation. The overall results globally show that the non-convex optimization provides better results in practice, although it does not guaranty obtaining the global minimizer. Finally, a general caveat was identified: the spatial sampling should be sufficiently dense, otherwise the covariance matrix is such that its off-diagonal elements have too small amplitudes to obtain a proper estimation of the gain phases. \added{The authors believe that this work can be extended at will to very large arrays as long as the sensor spatial density respects the over-sampling condition. In perspective, further realistic scenarios can be studied with unaddressed model mismatches, such as uncertain sensor positions on the array structure.}

\section{Acknowledgements}
We thank Johan Clavier and Thomas Rougier for their quality work in the realization of experiments. We also thank Cédric Herzet for the fruitful discussions and suggested directions concerning proximal methods. This study has been produced in the framework of LUG2 supported by Région Auvergne Rhône-Alpes and BPIFrance (FUI22). It was performed within the framework of the Labex CeLyA of Université de Lyon, operated by the French National Research Agency (ANR-10-LABX-0060/ANR-11-IDEX-0007).

\bibliography{sampbib}
\end{document}